\documentclass{article}
\usepackage{graphicx} 
\usepackage{booktabs}
\usepackage{amsmath}
\usepackage{geometry} 
\usepackage{hyperref}

\title{FPGA-Accelerated SpeckleNN with SNL for
Real-time X-ray Single-Particle Imaging}
\author{
    Abhilasha Dave\textsuperscript{1}, Cong Wang\textsuperscript{1}, James Russell\textsuperscript{1},  Ryan Herbst\textsuperscript{1}, and Jana Thayer\textsuperscript{1}\\
    \textsuperscript{1}SLAC National Accelerator Laboratory, Menlo Park, CA, USA \\
    \text{\{adave, cwang31, russell, rherbst, jana\}@slac.stanford.edu}
    }

\date{} 

\begin{document}
\maketitle

\section{Abstract}
\label{sec:abstract}

We present the implementation of a specialized version of our previously published unified embedding model, SpeckleNN, for real-time speckle pattern classification in X-ray Single-Particle Imaging (SPI), using the SLAC Neural Network Library (SNL) on an FPGA platform. This hardware realization transitions SpeckleNN from a prototypic model into a practical edge solution, optimized for running inference near the detector in high-throughput X-ray free-electron laser (XFEL) facilities, such as those found at the Linac Coherent Light Source (LCLS).

To address the resource constraints inherent in FPGAs, we developed a more specialized version of SpeckleNN. The original model, which was designed for broader classification across multiple biological samples, comprised approximately 5.6 million parameters. The new implementation, while reducing the parameter count to 64.6K (a 98.8\% reduction), focuses on maintaining the model's essential functionality for real-time operation, achieving an accuracy of 90\%. Furthermore, we compressed the latent space from 128 to 50 dimensions.

This implementation was demonstrated on the KCU1500 FPGA board, utilizing 71\% of available DSPs, 75\% of LUTs, and 48\% of FFs, with an average power consumption of 9.4W according to the Vivado post-implementation report. The FPGA performed inference on a single image with a latency of 45.015 microseconds at a 200 MHz clock rate.

In comparison, running the same inference on an NVIDIA A100 GPU resulted in an average power consumption of approximately 73W and an image processing latency of around 400 microseconds. Our FPGA-accelerated version of SpeckleNN demonstrated significant improvements, achieving an 8.9x speedup and a 7.8x reduction in power consumption compared to the GPU implementation.

Key advancements include model specialization and dynamic weight loading through SNL, which eliminates the need for time-consuming FPGA design re-synthesis, allowing fast and continuous deployment of models (re)trained online. These innovations enable real-time adaptive classification and efficient vetoing of speckle patterns, making SpeckleNN more suited for deployment in XFEL facilities. This implementation has the potential to significantly accelerate SPI experiments and enhance adaptability to evolving experimental conditions.

\section{Introduction}
\label{sec:introduction}

The rapid advancement of next-generation detectors in scientific research and industrial applications has catalyzed an exponential increase in data generation rates. Ultra-high-rate (UHR) detectors, such as those used in X-ray free-electron laser (XFEL) facilities like Linac Coherent Light Source (LCLS), now operate at frequencies exceeding 100 kHz, producing data throughputs that can surpass 1 TB/s. These cutting-edge technologies have revolutionized the study of nanoscale particles, particularly through X-ray Single-Particle Imaging (SPI). In these experiments, intense femtosecond X-ray pulses generate intricate scattering patterns, commonly referred to as ``speckles", from individual particles. These speckles are critical for reconstructing the three-dimensional structures of non-crystalline particles at room temperature. However, the high data rates at these facilities present significant challenges for real-time classification of speckle patterns, which is essential for accurate and rapid identification of single hits required for subsequent three-dimensional reconstructions.

Traditional approaches to speckle pattern classification have relied heavily on either unsupervised learning techniques, which often necessitate post-experimental human intervention, or supervised learning models that require extensive labeled datasets. These datasets are time-consuming to generate and are often impractical to obtain at the scale required for real-time applications in high-throughput environments. Consequently, both approaches are suboptimal for the rapid, on-the-fly analysis needed in XFEL facilities, where computational complexity and latency can significantly hinder experimental progress.

To address these challenges, we previously introduced SpeckleNN~\cite{wang2023specklenn}, a unified embedding model specifically designed for the real-time classification of speckle patterns with limited labeled examples. SpeckleNN employs a contrastive learning approach using twin neural networks \cite{bromley1993signature,chopra2005learning}, which map speckle patterns to a unified embedding vector space. Within this space, classification is performed based on the Euclidean distance between embedding vectors, enabling robust few-shot classification even in scenarios with sparse labeling or missing detector areas. This model is particularly suited for SPI experiments, where rapid and accurate classification is critical to the success of data collection and subsequent analysis.

Despite its effectiveness, maintaining ML inference at high data rate presented significant challenges.  While GPUs are powerful tools for parallel processing, their performance in real-time, edge deployments can be influenced by various factors, such as data transfer overhead, memory constraints, and the need for batch processing to optimize throughput. This is particularly problematic when data must be continuously streamed from sensors or detectors, and immediate feedback is required.

To overcome the challenge of real-time inference in high-throughput environments, there is a growing interest in computing platforms that can provide low-latency processing at the edge, in close proximity to the instrumentation. Field-Programmable Gate Arrays (FPGAs) have emerged as a promising solution, leveraging their advanced parallelism and configurability. FPGAs can be optimized for processing individual data points with minimal latency, making them particularly suitable for applications that require immediate feedback and real-time analysis. This capability is particularly advantageous in high data rate XFEL facilities.

However, the promise of FPGAs comes with its own set of challenges, particularly related to its resource constraints. These devices, while powerful, are typically smaller and offer limited computational resources compared to platforms like GPUs. This presents a significant hurdle when deploying neural network models, which often contain millions of parameters. Our research confronted this challenge with the original SpeckleNN model, which comprised nearly 5.5 million parameters. Fitting such a large model into single FPGA without exceeding its capacity was not feasible.

To address this, we undertook a rigorous optimization process that successfully reduced the model size by approximately 98.8\%. However, this reduction came at a cost to accuracy. While the original model achieved around 98\% accuracy, the newer, lighter version has an accuracy of 90\%. Trading off model size with accuracy in this case is appropriate, enabling the SpeckleNN model to run efficiently on edge devices like FPGAs, thus facilitating real-time data processing directly at the site of data collection. Despite the challenges posed by the reduction in model size, the benefits of deploying these optimized models on FPGAs are profound. By enabling inference on a per-image basis with substantially reduced latency, FPGAs eliminate the bottlenecks associated with batch size constraints in GPUs, allowing for continuous, real-time data analysis without the burdensome overhead of data transfer and storage.

To deploy this optimized model on FPGAs, we leveraged the SLAC Neural Network Library (SNL)~\cite{herbst2022implementation}, a powerful tool designed specifically to address the challenges of high-throughput, low-latency environments. SNL enables the seamless translation of machine learning architectures into FPGA-compatible code, allowing for the construction of data processing pipelines characterized by ultra-low latency and high throughput. This capability is particularly crucial for managing the immense data velocities generated by modern detectors.

One of the most significant advantages of SNL is its ability to dynamically load weights and biases onto the FPGA. This feature eliminates the need for resynthesizing the entire neural network code for FPGA when the model is retrained, enabling immediate inference runs with updated models. This feature greatly enhances the flexibility and efficiency of using FPGAs in environments where models may need to be frequently updated or fine-tuned to adapt to new data or experimental conditions. As a result, deploying machine learning models on FPGAs via SNL represents a significant advancement, offering a robust solution for real-time data processing across a wide range of scientific domains.

This paper details the implementation of a streamlined version of SpeckleNN on the KCU1500 FPGA board using Xilinx Vitis ~\cite{kathail2020xilinx} and the SLAC Neural Network Library (SNL) ~\cite{xilinx_kcu1500}. This approach demonstrates the board's capability to meet the stringent demands of high-throughput, low-latency environments. Key advancements include extensive model pruning and the novel application of dynamic weight loading through SNL, eliminating the need for FPGA re-synthesis and enabling online retraining for continuous model improvement. These innovations enable real-time adaptive classification and efficient vetoing of speckle patterns, optimizing SpeckleNN for deployment in XFEL facilities. Consequently, this implementation accelerates SPI experiments and enhances system adaptability to evolving experimental conditions.

The primary objective of this study is to demonstrate the feasibility and effectiveness of deploying a highly optimized version of the SpeckleNN model on Field-Programmable Gate Arrays (FPGAs) for real-time speckle pattern classification in X-ray single-particle imaging (SPI) at X-ray free-electron laser (XFEL) facilities. Specifically, the study aims to:
\begin{enumerate}
     \item \textbf{Optimize the SpeckleNN Model:}  Achieve a significant reduction in the model size, from ~5.6 million parameters to ~64.6K parameters (~98.8\% reduction), while maintaining high classification accuracy (~90\%), making it suitable for deployment on resource-constrained FPGA devices.
    \item \textbf{Implement and Evaluate on FPGA:} Implement the optimized SpeckleNN model on the KCU1500 FPGA board using the SLAC Neural Network Library (SNL) and evaluate its performance in terms of resource utilization (DSPs, LUTs, FFs), power consumption, and inference latency.
    \item \textbf{Compare FPGA Performance to GPU:} Compare the FPGA-accelerated SpeckleNN’s performance with that of a GPU (NVIDIA A100), focusing on improvements in speed (latency reduction) and power efficiency.
    \item \textbf{Consider Arithmetic Precision:} Conduct a detailed consideration of arithmetic precision for each layer of the FPGA-based implementation, comparing the outcomes of FPGA-based C-Simulation (CSIM) with those from PyTorch-based simulations. This ensures that the precision and accuracy of the model are maintained across different computational environments, which is crucial for reliable real-time inference.
\end{enumerate}

\section{Methodology}
\label{sec:methodology}
The methodology for this study is structured around four key objectives mentioned in the introduction section earlier, each aimed at ensuring the successful deployment of the SpeckleNN model on FPGA hardware while maintaining performance, efficiency, and accuracy.
\begin{figure}
    \centering
    \includegraphics[width=0.85\linewidth]{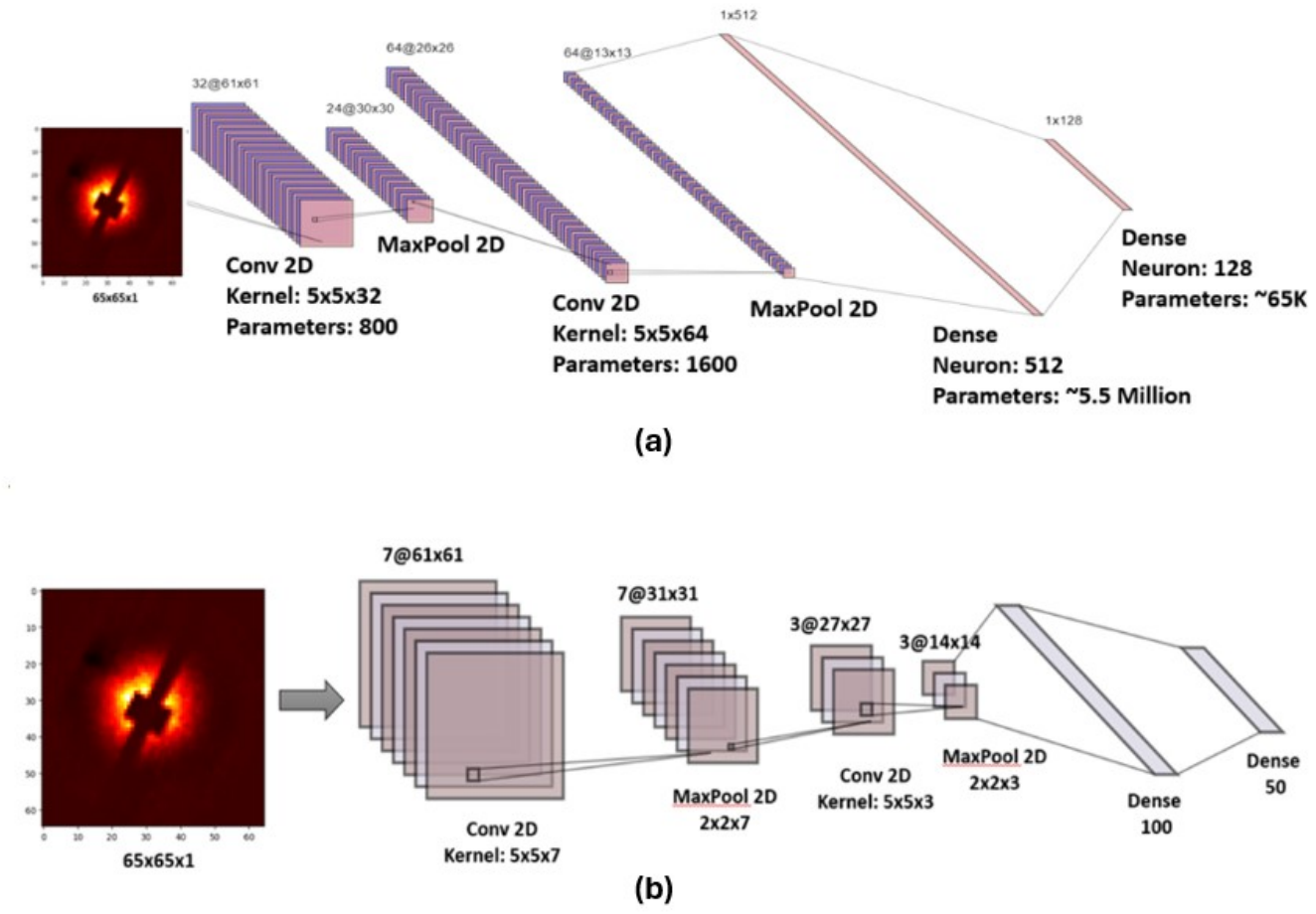}
    \caption{SpeckleNN model (a) original model architecture (b) optimized model architecture}
    \label{fig:SpeckleNN}
\end{figure}

\subsection{SpeckleNN Model Architecture and Optimization}

To optimize our neural network for deployment on resource-constrained devices such as FPGAs, we systematically reduced the model size while rigorously evaluating its impact on classification accuracy. The architecture of both the original and optimized models used in our experiments is illustrated in Figure 1.

\begin{itemize}
    \item \textbf{Original Model Architecture} The original SpeckleNN model (Figure 1(a)) featured two convolutional layers followed by two dense layers, which, while delivering high accuracy, also resulted in a computationally intensive design. Specifically, the model's architecture included 32 filters of size 5x5 in the first convolutional layer, paired with a 2x2 max-pooling layer. The second convolutional layer expanded to 64 filters, again followed by a max-pooling layer. These convolutional layers fed into a dense layer with 512 units, culminating in a final dense layer of 128 units. This design achieved a classification accuracy of 98\%, but the model's approximately 5.6 million parameters posed significant challenges for deployment on FPGAs due to their constrained computational resources.

    \item \textbf{Optimized Model Architecture} Recognizing the necessity for a more resource-efficient design, we optimized the SpeckleNN architecture with a focus on minimizing the parameter count while maintaining high accuracy. The optimized model (Figure 1(b)) incorporated several modifications: the first convolutional layer was reduced to 7 filters of size 5x5, followed by a 2x2 max-pooling layer, and the second convolutional layer was minimized to 3 filters, again with max-pooling. The dense layers were similarly scaled down to 100 and 50 units, respectively. Through this process, we achieved a dramatic reduction of 98.8\% in the number of parameters—from 5.6 million to approximately 64.6K—while maintaining a robust classification accuracy of 90\%. This demonstrates that the original larger model, with its 5.6 million parameters, had redundancy that wasn’t necessary for effective learning. By using fewer parameters, the optimized model has become more efficient, with each parameter contributing effectively to the network's final accuracy. This compact architecture ensures that the model not only retains strong predictive performance but also operates more efficiently, making it ideal for deployment on resource-constrained FPGAs where rapid, real-time processing and resource efficiency are paramount.

    \item \textbf{32-bit Inference Implementation} We implemented the inference run using 32-bit floating-point precision to strike a balance between computational efficiency and maintaining high accuracy. Although lower-bit precision could further reduce resource usage and power consumption, it often introduces quantization errors that can degrade accuracy, especially in deep neural networks \cite{pytorch_numerical_accuracy, codesign_BNN}. Given the dramatic reduction in model size, where we achieved 90\% accuracy, our priority was to preserve this accuracy rather than sacrifice it by aggressively quantizing the parameters \cite{quant_acc}. Since the 32-bit version fit successfully on the KCU1500 FPGA board, we opted to keep this precision for now. However, in future studies, we plan to explore the potential of parameter quantization to further optimize the model's performance without compromising accuracy.

    \item \textbf{Trade-off Analysis} The optimization process inevitably involved a trade-off between model size and accuracy. The reduction in parameters led to a slight decrease in accuracy from 98\% to 90\%. However, this trade-off is often justified in scenarios where deployment constraints, such as limited computational resources or the necessity for real-time processing, are paramount. The significant reduction in model size not only accelerates inference times but also substantially reduces the computational load and memory footprint, making the optimized model more suitable for high-throughput environments like those encountered in XFEL facilities. 
\end{itemize}

In summary, the optimized SpeckleNN model effectively balances accuracy, efficiency, and precision. By reducing the parameter count, we enhanced the effective use of each parameter towards achieving high accuracy, and by implementing a 32-bit inference run, we ensured that the model operates efficiently while maintaining its performance integrity. These optimizations make the model an ideal candidate for deployment in edge computing environments, where rapid and efficient data processing is essential.

\subsection{Implementation and Evaluation on FPGA using SNL}

After optimizing the SpeckleNN model, as shown in Figure 1(b), we moved forward with its implementation on the FPGA platform using the SLAC Neural Network Library (SNL). As illustrated in Figure 2, we began by defining the architecture of the SpeckleNN neural network using PyTorch \cite{pytorch}, a widely-used machine learning framework. The PyTorch model, which provides a high-level and flexible environment for neural network design, was translated into a C++ parameter template within the Xilinx Vitis platform \cite{kathail2020xilinx, amd_vitis_hls_templates} to facilitate its deployment using the SLAC Neural Network Library (SNL). SNL, designed specifically for FPGA-based machine learning implementations, operates using a C++ template model where weights, and biases are represented as memory-mapped interfaces. This allows for the precise mapping of neural network parameters to FPGA hardware, ensuring that the network structure is preserved while optimizing for performance in resource-constrained environments.

In SNL, the memory-based interface manages the storage and retrieval of weights and biases, allowing the FPGA to access these parameters efficiently during inference. The communication between different layers of the neural network is handled through a streaming interface, where data flows sequentially from one layer to the next without the need for intermediate storage. This streaming interface is particularly advantageous for real-time processing as it minimizes latency and maximizes throughput, both of which are critical for high-performance edge applications like those found in XFEL facilities.

The use of a C++ parameter template in SNL also supports a modular and re-configurable design, which is crucial when implementing complex neural networks such as SpeckleNN on hardware. By directly mapping the PyTorch model into this C++ structure, we ensured that the translation from software to hardware was seamless, preserving the integrity of the model’s architecture while making it compatible with FPGA constraints. 

Furthermore, SNL’s design allows for a clear separation between the computational elements (such as matrix multiplications and convolutions) and the memory management for storing weights and biases. This separation ensures that the FPGA’s computational resources are used efficiently, with minimal contention for memory access during the inference process. The streaming interface between layers further reduces bottlenecks, allowing data to flow continuously through the pipeline, thus achieving low-latency performance. This structure enables real-time inference on large datasets without sacrificing accuracy or speed, making it ideal for high-throughput, low-latency environments.

Overall, by leveraging SNL’s C++ parameter template and streaming architecture, we successfully implemented the optimized SpeckleNN model on FPGA hardware. This approach allowed us to maintain the integrity of the PyTorch-based design with a reasonable 90\% accuracy, while ensuring that the model was adapted to meet the performance demands of FPGA deployment, specifically in scientific applications requiring real-time SPI data processing in XFEL facilities.

\begin{figure}
    \centering
    \includegraphics[width=0.8\linewidth]{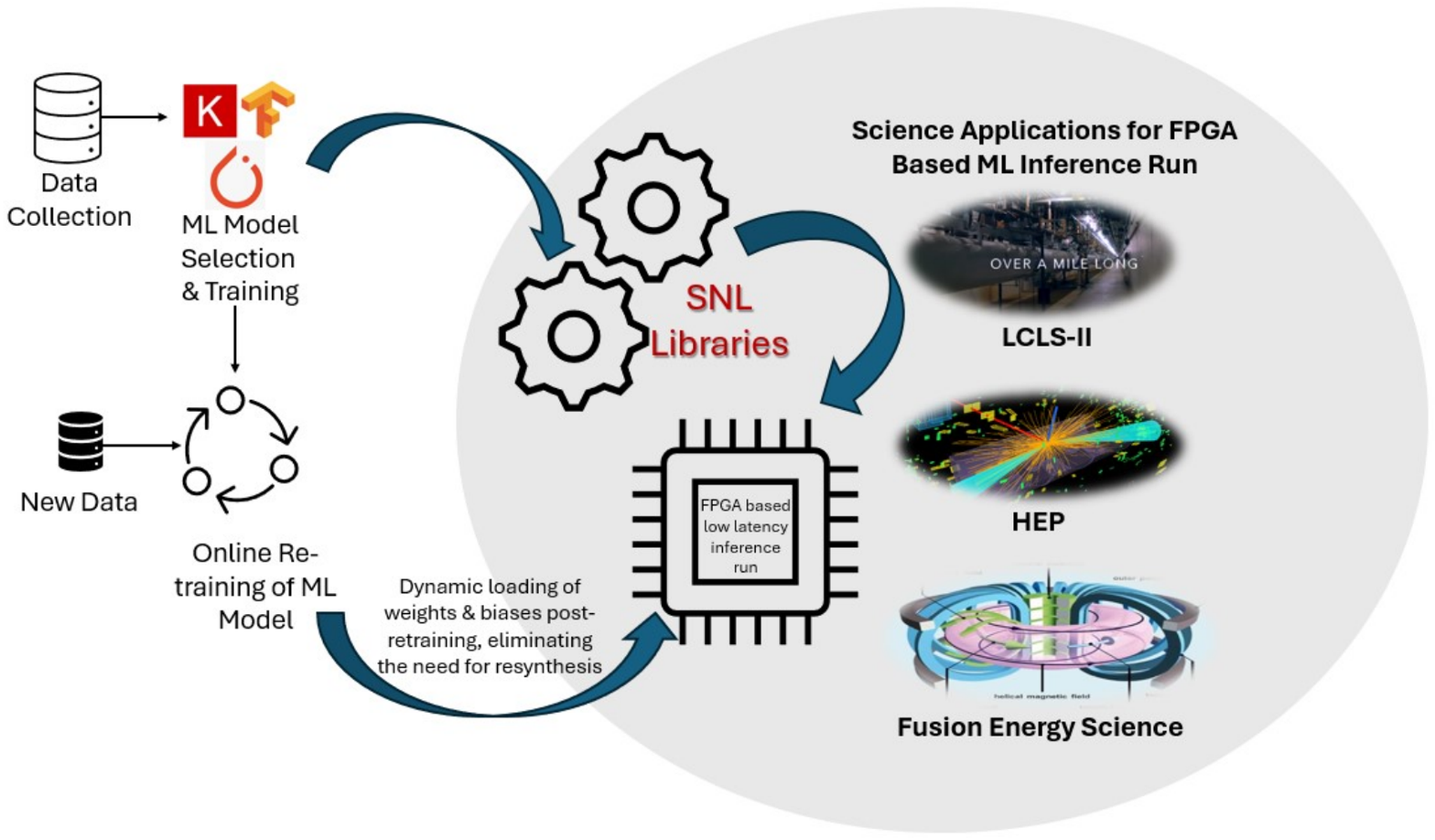} 
    \caption{SNL High Level Design Flow}
    \label{fig:SNL-LOGO}
\end{figure}

\subsubsection{Evaluation of each layer type Outcomes Between SNL-Csim and PyTorch}

In this section, we conducted a comprehensive evaluation by comparing the output of various layers in the SpeckleNN model across two platforms: SNL’s C-simulation (Csim) and PyTorch. The comparison includes a single layer of each type—convolution, ReLU activation, MaxPool, and the final two dense layers to examine the behavior of these layers in both frameworks. While both Csim and PyTorch are designed for deep learning model implementation, they can differ in how they handle numerical precision and arithmetic operations, potentially leading to slight discrepancies in results. These differences primarily stem from variations in floating-point arithmetic, internal optimizations, and the way each framework manages edge cases such as overflow, underflow, and rounding.\\
Machine learning frameworks like PyTorch and Keras ~\cite{northcutt2021reproducibility, northcutt_benchmarking,viso2024pytorch,pytorch_forum2017,towardsdatascience2019kerasvspytorch} are known to produce slightly different results even when performing the same operations. These differences stem from how they manage floating-point precision and internal optimizations ~\cite{johnson2018rethinking,fpu,NNTraining_limitedPrecision}. While such discrepancies may seem minor in the early layers of a neural network, they can accumulate and propagate through subsequent layers \cite{codesign_BNN}, potentially affecting the final model output. Therefore, a layer-by-layer comparison is crucial to understand how these numerical differences affect the SpeckleNN model's behavior when deployed on an FPGA using SNL compared to the PyTorch reference.

\textbf{Layer0 Convolutional Feature Maps analysis (Pre-ReLU activation):}

As shown in Figure 3 we compared the outputs from Csim and PyTorch for Layer 0 across seven channels, immediately following the convolution operation. The key observations were:
\begin{figure}
    \centering
    \includegraphics[width=1\linewidth]{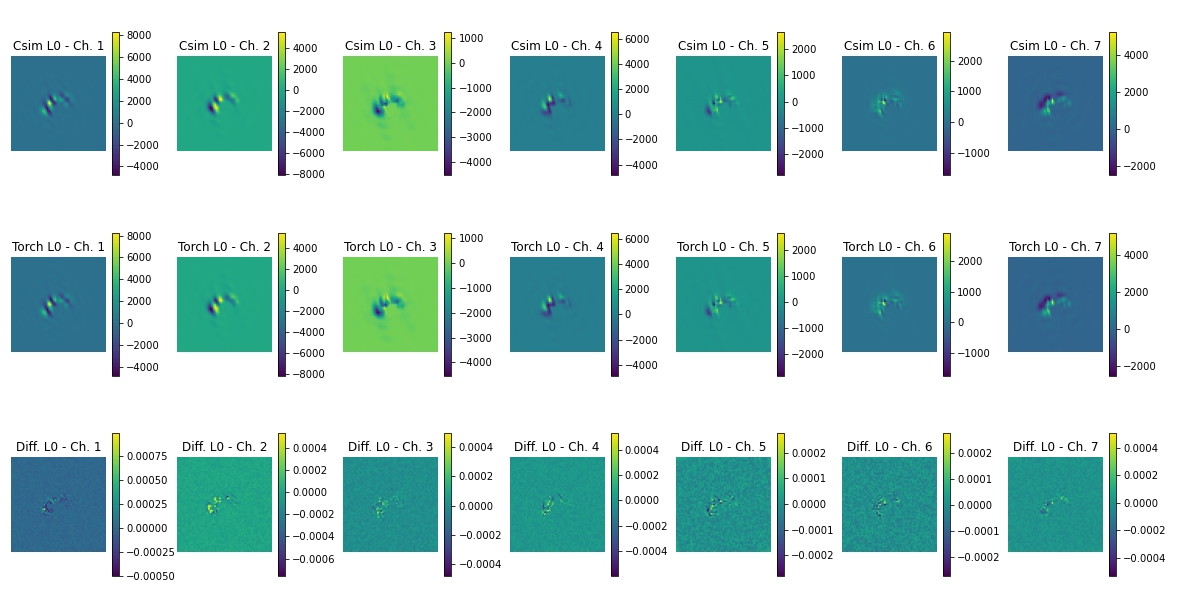}
    \caption{Convolutional Layer 0 outcome of all 7 output featuremap of SNL Csim, PyTotch, and difference between SNL Csim and PyTorch}
    \label{fig:Convo0_Output_Featuremap}
\end{figure}
\begin{itemize}
    \item High Consistency: Both frameworks produced nearly identical feature maps across all channels. Minor differences were observed, most notably in Channel 3, where the difference between the two frameworks was slightly larger but still within a minimal range (around -0.0004 to +0.0004).
    \item Minor Numerical Differences: The pixel wise differences were very small, mostly due to floating-point precision differences, with the majority of the differences falling within the range of -0.0005 to 0.0005, indicating that the two frameworks are well-aligned in their convolution computation.
\end{itemize}

\textbf{Layer0 ReLU Activation Applied to the Convolutional Feature Maps:}

Figure 4 presents the comparison of the resulting feature maps from both frameworks after applying the ReLU activation function to the convolutional outputs. The key findings are as follows:
\begin{figure}
    \centering
    \includegraphics[width=1\linewidth]{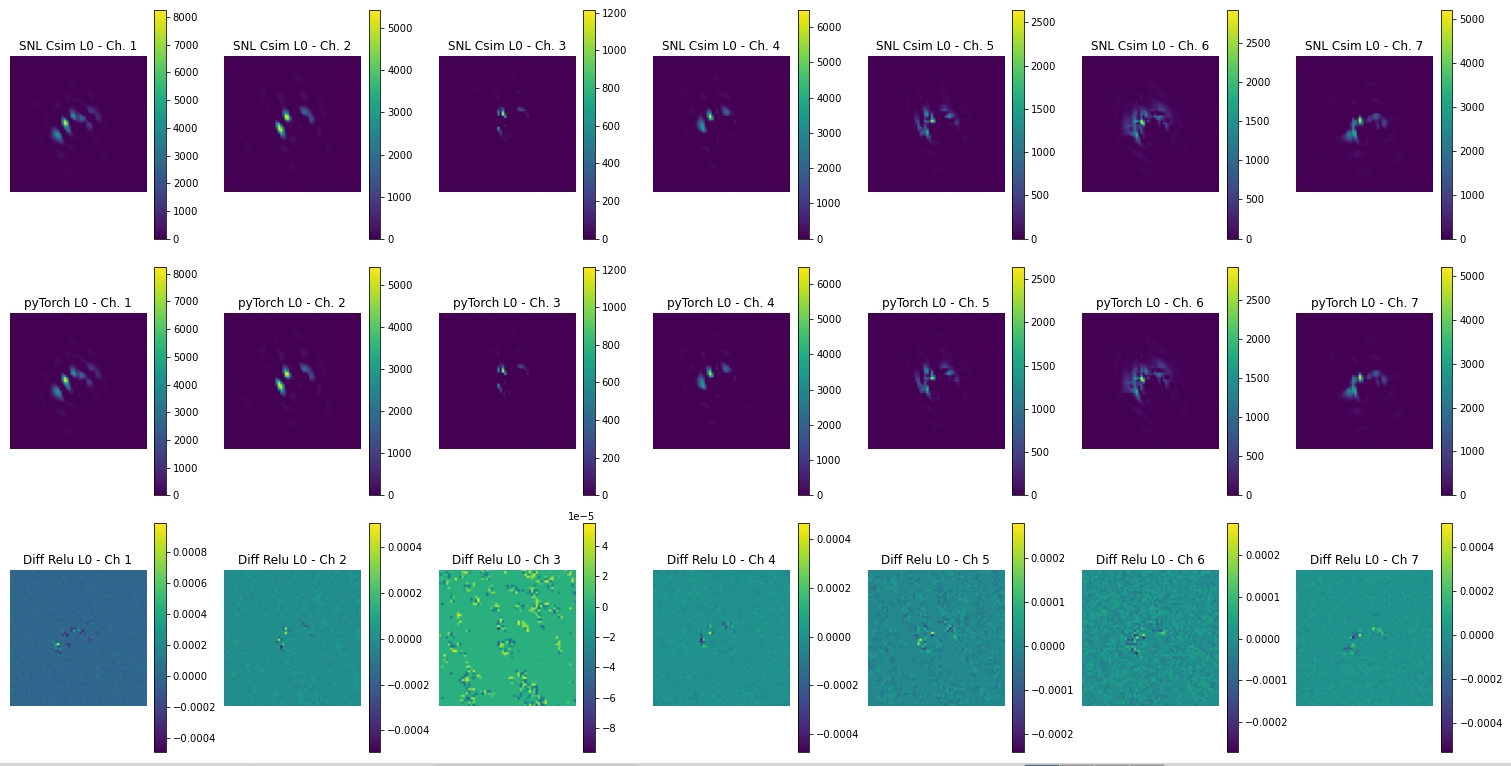}
    \caption{Convolutional Layer 0 outcome after passing through ReLU activation for all 7 output featuremap of SNL Csim, PyTotch, and difference between SNL Csim and PyTorch}
    \label{fig:Convo0_ReluOutput}
\end{figure}
\begin{itemize}
    \item Sparse Activations: As expected, ReLU zeroed out all negative values in the feature maps, resulting in sparse activations, where only positive values remained.
    \item Consistency maintained: The ReLU-activated feature maps remained highly consistent between Csim and PyTorch across all channels. The visual similarity between the frameworks was nearly identical, reflecting the same activation patterns.
    \item Channel 3 Differences: Although differences remained minimal, Channel 3 again showed slightly larger deviations compared to the other channels. These differences, in the range of $-4\times10^{-5}$ to $4\times10^{-5}$, were more noticeable than in other channels, but still very small and insignificant in terms of model performance.
\end{itemize}
\textbf{Connecting the Two Observations for layer 0 before and after ReLU:}
\begin{itemize}
    \item ReLU's Impact on Differences: The differences we observed in the convolutional outputs (particularly in Channel 3) persisted post-ReLU, though they remained very small. ReLU did not amplify these differences, suggesting that the inconsistencies seen in Channel 3 are inherent to the convolutional computation itself, rather than being introduced by ReLU activation.
    \item Consistency Across Layers: The high consistency between Csim and PyTorch in both the convolutional output and post-ReLU activation maps indicates that both frameworks handle the convolution and activation operations similarly, with only minor numerical variations. These variations are within an acceptable range and do not suggest any significant divergence in the implementation of these operations.
    \item The convolutional outputs of Csim and PyTorch are highly consistent, with only minor numerical differences that carry through after the application of ReLU. These small differences, particularly in Channel 3, suggest that some channels may be more sensitive to precision, but these deviations are unlikely to affect overall model performance. 
\end{itemize}

\textbf{Analysis and Connection Between Convolutional and Dense Layer Outputs}
\begin{figure}
    \centering
    \includegraphics[width=1\linewidth]{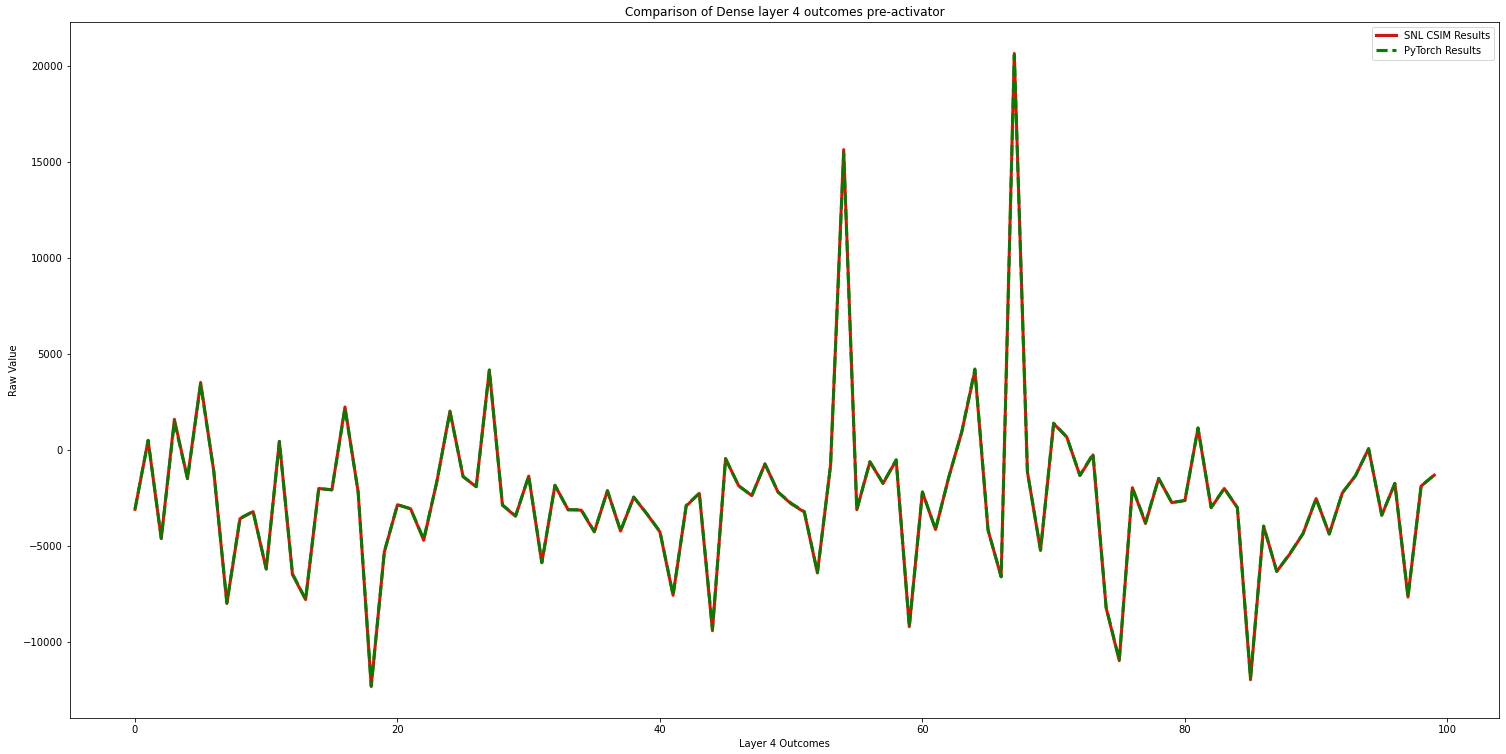}
    \caption{Dense layer 4 output SNL Csim and PyTorch}
    \label{fig:Dense0_output_featuremap}
\end{figure}

\begin{figure}
    \centering
    \includegraphics[width=1\linewidth]{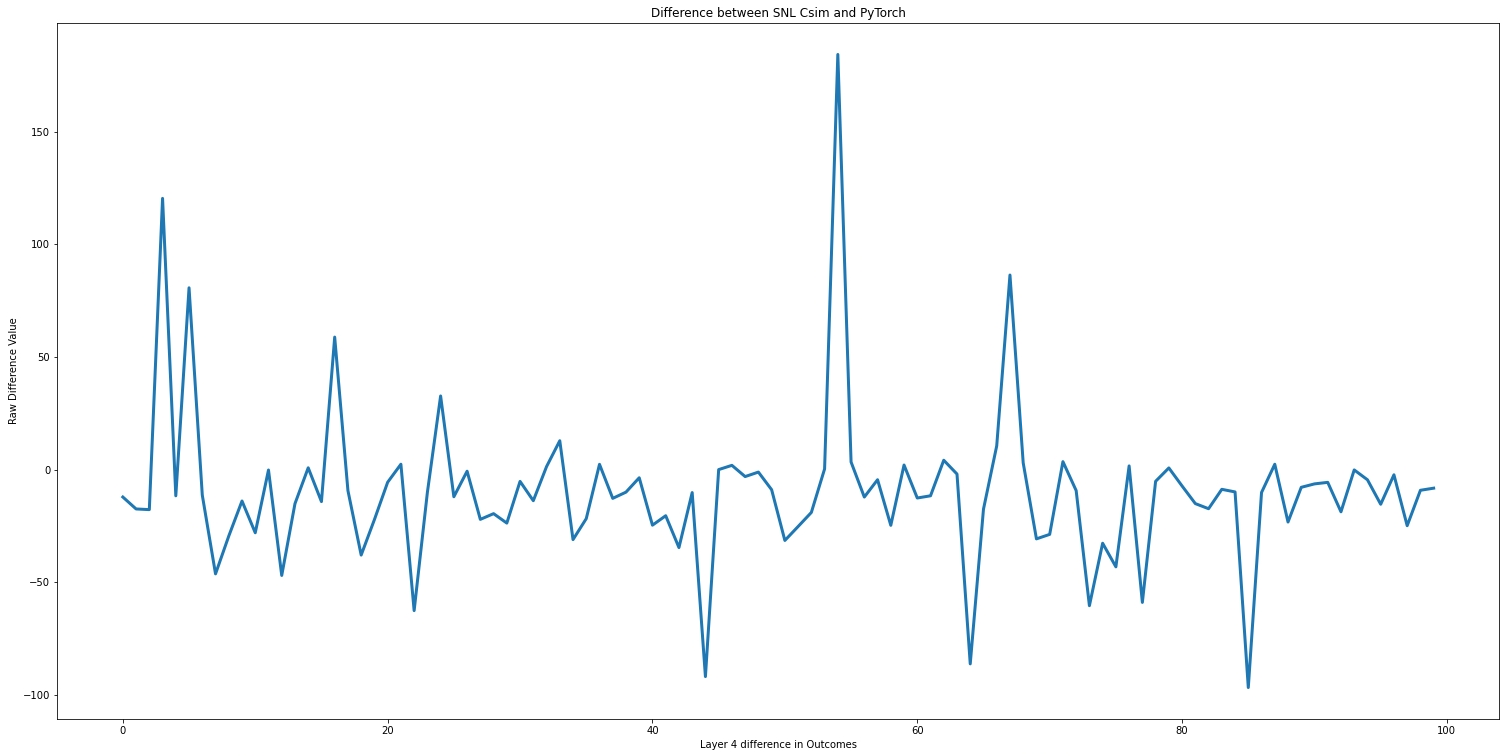}
    \caption{Dense layer 4 output difference between SNL Csim and PyTorch}
    \label{fig:Dense0_outputDiff_featuremap}
\end{figure}

\begin{itemize}
    \item In our previous analysis of layer 0, we first observed the convolutional output feature maps, which revealed that the major activity or significant patterns were concentrated in the central regions of the feature maps across multiple channels. This indicates that the convolutional layers capture prominent features, particularly in the middle of the input space, where activations are more pronounced.
    \item When these activations are propagated through the network, they have a direct influence on the dense layer outputs, as observed in the subsequent analysis. As shown in Figure 5 the comparison of dense layer (Layer 4) before ReLU activation outputs from SNL Csim and PyTorch shows that both frameworks exhibit similar neuron activation patterns, with the most significant spikes in neuron values occurring around neurons 60 to 70. These spikes can be attributed to the feature extraction process of the convolutional layer, where the highly active regions in the middle of the feature maps are likely to contribute to stronger connections to specific neurons in the dense layer. The dense layer is designed to aggregate the most relevant information from the convolutional layers, and the prominent spikes observed in neurons 60 to 70 are a direct reflection of the amplified signals derived from the high-activation areas in the convolutional outputs.
    \item Further comparison between SNL Csim and PyTorch reveals high consistency between the two frameworks, as shown in Figure 5 by the overlapping plots in the pre-activation outputs. However, the small numerical differences, particularly the localized spikes in differences around neurons 10, 30, and 60 shown in the difference plot figure 6, likely come from floating-point precision variations or minor differences in internal precision handling between the two frameworks. These discrepancies, while noticeable in certain neurons, remain relatively small compared to the overall magnitude of the neuron outcomes, which range from approximately -20,000 to +20,000.
    \item The analysis demonstrates that the significant spikes in neuron activation in the dense layer are a direct consequence of the prominent features extracted by the convolutional layers, particularly in the middle of the feature maps. These high-activation regions are captured and amplified by the dense layer neurons, leading to larger spikes in output. Despite small numerical differences between SNL Csim and PyTorch, both frameworks perform consistently, and the minor deviations observed are not expected to significantly impact overall model performance. This consistency reinforces the robustness of both frameworks in implementing deep learning models with comparable behavior.
\end{itemize}
\textbf{Implications and Key Takeaways}
\begin{enumerate}
    \item High Framework Consistency: The comparison between SNL Csim and PyTorch reveals strong consistency, with only minor numerical differences. These differences, observed around specific neurons (10, 30, and 60), are likely due to floating-point precision variations and framework-specific optimizations. However, these discrepancies are minimal and do not significantly affect the overall model behavior, ensuring reliability in implementation across both platforms.
    \item Convolutional Features Driving Dense Layer Activations: The spikes observed in the dense layer outputs, particularly around neurons 60 to 70, align with the high-activation regions in the convolutional feature maps. This shows that the convolutional layers effectively extract important features, which are then amplified by the dense layers. This insight reinforces the importance of optimizing the feature extraction process in convolutional layers to improve downstream activations in dense layers.
    \item Framework Interoperability and Model Robustness: The high degree of similarity between the two frameworks demonstrates their interoperability, allowing models to be trained in one and deployed in the other without significant performance loss. This flexibility is crucial for model development and deployment across different platforms. Furthermore, the model’s robustness to small numerical variations indicates that it can produce stable results across frameworks, enhancing its applicability in real-world scenarios.
    \item Handling Arithmetic Discrepancies Across Platforms: Despite these differences, the error introduced by the arithmetic handling between SNL-Csim and PyTorch remains within acceptable limits, allowing the model to maintain comparable accuracy between FPGA-based inference and PyTorch simulations. For the SpeckleNN model, the accumulated error is still small enough that it does not significantly impact final accuracy, as demonstrated by the near-identical results between the two platforms. However, this evaluation raises an important question: as models transition between different frameworks and hardware platforms, how can we ensure consistent handling of arithmetic operations? Small discrepancies, especially in deeper networks, could accumulate and affect the reliability of models across platforms. Addressing these concerns is critical for ensuring robustness and consistency when moving from software environments like PyTorch to hardware-based implementations, such as FPGAs via SNL-Csim.

    \item Need for Further Exploration of Numerical Precision: This analysis highlights the need for further exploration into how numerical precision and arithmetic operations are handled across platforms. In edge applications, where minor discrepancies could have significant impact on model performance, ensuring consistency becomes even more crucial. Future research should focus on addressing these precision-related issues to improve the reliability and robustness of machine learning models when transitioning between software and hardware platforms.
\end{enumerate}

\subsection{Resource Utilization and Performance Analysis on KCU1500 FPGA}

\textbf{Resource Utilization and Performance Evaluation on the KCU1500 FPGA Board}

After successfully implementing the SpeckleNN network on the KCU1500 FPGA board \cite{xilinx_kcu1500} , we conducted a detailed evaluation of the resource utilization and inference performance. The results, as depicted in Figure 7, demonstrate the efficient use of FPGA resources. Specifically, the implementation utilized 75\% of the available Look-Up Tables (LUTs), 48\% of Flip-Flops (FFs), 25\% of Block RAM (BRAM), and 71\% of Digital Signal Processors (DSPs). This balanced utilization indicates that the network design was well-optimized to fit within the resource constraints of the KCU1500 board.

\begin{figure}[ht]
    \centering
    \includegraphics[width=1\linewidth]{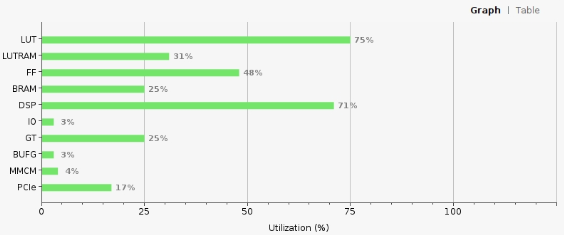}
    \caption{KCU1500 FPGA Board Resource Utilization}
    \label{fig:FPGA-Resource-Utilization00}
\end{figure}

\textbf{Inference Performance and Latency}
For the inference run of the neural network on the FPGA, as shown in Figure 8, we employed the Integrated Logic Analyzer (ILA) for debugging and performance monitoring. The ILA captured the complete frame of the inference process, tracking the time from the arrival of the first pixel to the generation of the final 50th dimensional latent space output. The total time for this operation was approximately 9,003 FPGA clock cycles. Given the clock cycle rate of 5 nanoseconds, the total latency for the inference run was measured at approximately 45.05 microseconds (µs).

\begin{figure}[ht]
    \centering
    \includegraphics[width=1\linewidth]{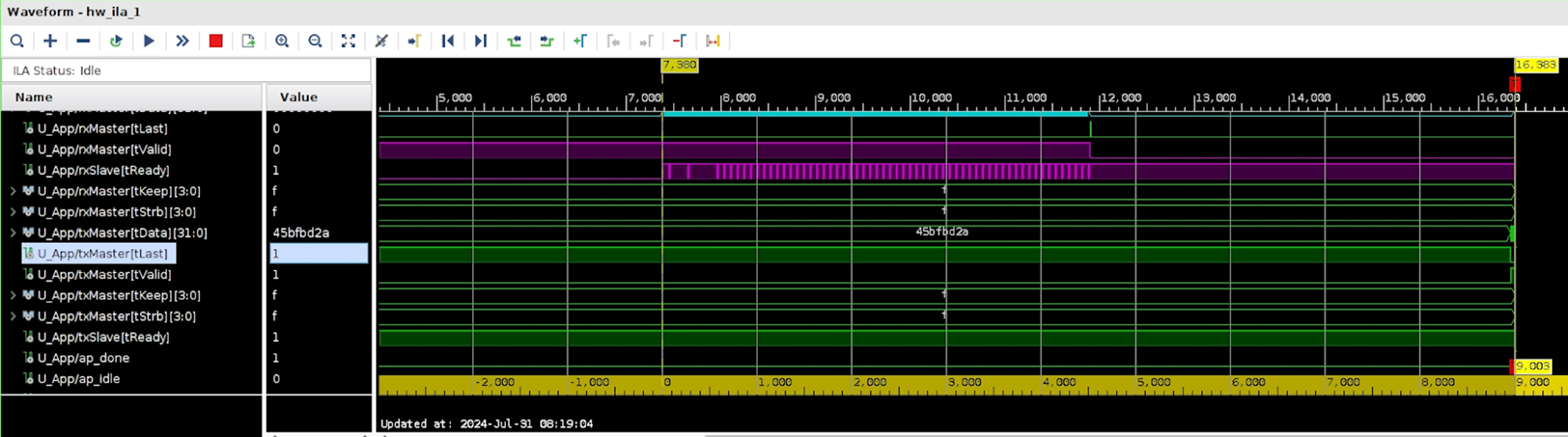}
    \caption{FPGA ILA Debug Latency Report}
    \label{fig:FPGA-ILA-Debug}
\end{figure}

This low-latency performance is critical for real-time applications in high-throughput environments, such as those found in XFEL facilities. The ability to process incoming data with such minimal delay ensures that the system can keep pace with the rapid data generation rates typically associated with these scientific experiments.

\textbf{Comparative Analysis with GPU: }

As highlighted in Table 1, the FPGA-based implementation of SpeckleNN provided significant improvements in both speed and power efficiency compared to the GPU-based inference run on the NVIDIA A100. Specifically, the FPGA achieved a remarkable 8.9x improvement in inference speed over the GPU, thanks to its lower latency and more efficient parallelism in handling the data. Additionally, the FPGA consumed 7.8x less power than the GPU, underscoring its suitability for edge computing environments where energy efficiency is critical.

These improvements demonstrate the effectiveness of deploying the optimized SpeckleNN model on FPGA hardware, not only in terms of computational speed but also in significantly reducing the power overhead, which is a key consideration for continuous, real-time processing in high-data-rate applications.

\begin{table}[ht]
    \centering
    \caption{Resource Utilization Analysis: SNL and HLS4ML for FPGA Inference Runs}
    \begin{tabular}{@{}lcc@{}}  
        \toprule
        \textbf{Inference Run Framework} & \textbf{Latency/image (µs)} & \textbf{Power (W)} \\
        \midrule
        SNL-based FPGA SpeckleNN & 45.05 (5ns clock period) & 9.4 \\
        GPU A100 & 400 & 73 \\
        \bottomrule
    \end{tabular}
    \label{tab:resource-utilization}
\end{table}

\section{Conclusion}
\label{sec:conclusion}
In this study, we successfully optimized and implemented the SpeckleNN neural network on the KCU1500 FPGA platform using the SLAC Neural Network Library (SNL) to meet the high-throughput, low-latency demands of real-time X-ray single-particle imaging (SPI) at X-ray free-electron laser (XFEL) facilities. The original SpeckleNN model, which initially consisted of 5.6 million parameters, was reduced by approximately 98.8\% to 64.6K parameters. This reduction, achieved through architectural modifications and parameter compression, allowed the model to fit efficiently on FPGA hardware while maintaining an impressive 90\% classification accuracy.

We chose to implement the network using 32-bit floating-point precision to balance computational efficiency and accuracy. While lower-bit precision could potentially reduce resource consumption further, the 32-bit precision provided a level of reliability and compatibility with industry-standard frameworks such as PyTorch. This decision also ensured that the FPGA implementation closely matched the performance and behavior of the original PyTorch model, minimizing discrepancies due to arithmetic handling differences.

Our evaluation of resource utilization showed that the SpeckleNN model efficiently utilized 75\% of LUTs, 48\% of Flip-Flops, 25\% of BRAM, and 71\% of DSPs on the KCU1500 board, confirming that the design was well-optimized for the FPGA’s resource constraints. The inference latency was measured at 45.05 microseconds, with a total of 9,003 clock cycles, meeting the real-time processing requirements essential for high-throughput environments.

A layer-by-layer comparison of the SNL C-simulation (Csim) results with PyTorch revealed small but manageable numerical discrepancies between the two platforms. While the error propagated through the deeper layers of the network, the difference remained within acceptable limits, ensuring that the FPGA’s performance closely mirrored the results obtained from the PyTorch simulations.

Finally, the comparative analysis between FPGA and GPU performance demonstrated significant advantages for FPGA-based deployment. The FPGA achieved a 8.9x speed improvement and consumed 7.8x less power compared to the NVIDIA A100 GPU, making it a highly efficient alternative for real-time, edge-based machine learning tasks.

In conclusion, our implementation of SpeckleNN on the KCU1500 FPGA, supported by SNL, proved to be a powerful and efficient solution for real-time speckle pattern classification in XFEL facilities. While we achieved substantial speed and power improvements, future work will explore further optimization techniques such as quantization to reduce resource usage while maintaining accuracy. This study underscores the potential of FPGA-based machine learning models for high-performance scientific applications, offering a compelling path forward for the deployment of neural networks in resource-constrained, real-time environments.

\section{Funding}
\label{sec:funding}
This material is based upon work supported by the U.S. Department of Energy, Office of Science, Office of Basic Energy Sciences under Award Number FWP-100643. Use of the Linac Coherent Light Source (LCLS), SLAC National Accelerator Laboratory, is supported by the US DOE, Office of Science, Office of Basic Energy Sciences (contract No. DE-AC02-76SF00515)

\section{Data Availability Statement}
\label{sec:data_availability_statement}
We obtained speckle data from an LCLS experiment studying bacteriophage PR772, collected at the AMO instrument (experiment ID amo06156; run numbers: 90, 91, 94, 96, and 102). The data can be found at \url{https://www.cxidb.org/id-156.html}.

\bibliographystyle{iopart-num}
\bibliography{main.bib}
\end{document}